\documentclass[%
twocolumn, prl, aps, superscriptaddress, nolongbibliography,showpacs,amsmath,amssymb,floatfix
]{revtex4-2}

\usepackage{graphicx}% Include figure files
\usepackage{dcolumn}% Align table columns on decimal point
\usepackage{bm}% bold math
\usepackage{graphicx}                                               
\usepackage{amssymb}

\usepackage{amsmath}
\usepackage{epsfig}
\usepackage{xcolor}
\usepackage{tabu}
\usepackage{mathtools}
\usepackage[colorlinks,linkcolor=blue,anchorcolor=blue,citecolor=blue,urlcolor=blue]{hyperref}
\usepackage{physics}
\usepackage{float}
\usepackage{diagbox}
\usepackage{inputenc}
\usepackage{pifont}
\usepackage{bbding}
\begin{document}

\preprint{APS/123-QED}

\title{Pairing-induced Momentum-space Magnetism and Its Implication In Optical Anomalous Hall Effect In Chiral Superconductors}
\author{Bin Geng}
\affiliation{CAS Key Laboratory of Strongly-Coupled Quantum Matter Physics, and Department of Physics, University of Science and Technology of China, Hefei, Anhui 230026, China}

\author{Yang Gao}
\email{ygao87@ustc.edu.cn}
\affiliation{CAS Key Laboratory of Strongly-Coupled Quantum Matter Physics, and Department of Physics, University of Science and Technology of China, Hefei, Anhui 230026, China}
\affiliation{ICQD, Hefei National Laboratory for Physical Sciences at Microscale, University of Science and Technology of China, Hefei, Anhui 230026, China}
 \affiliation{Hefei National Laboratory, University of Science and Technology of China, Hefei, Anhui 230026, China}

\author{Qian Niu}
\affiliation{CAS Key Laboratory of Strongly-Coupled Quantum Matter Physics, and Department of Physics, University of Science and Technology of China, Hefei, Anhui 230026, China}

\date{\today}% It is always \today, today,

\begin{abstract}
The intrinsic mechanisms of the magneto-optical Kerr signal in chiral superconductors often involve multi-orbital degree of freedom. Here by considering a generic single-orbital and spinful Hamiltonian, we generalize the Onsager's relation to obtain the necessary conditions for the optical anomalous Hall effect. Using the down-folding method, we identify two types of effective momentum-space magnetism responsible for the optical anomalous Hall conductivity from non-unitary and unitary pairing potentials respectively. The former is due to the angular momentum of Cooper pair, while the latter requires the participation of the spin-orbit coupling in the normal state and has been largely overlooked previously. Using concrete examples, we show that the unitary pairing can lead to both ferromagnetism and complicated antiferromagnetic spin texture in the momentum space, resulting in an in-plane optical anomalous Hall effect with the magnetism parallel to the Hall-deflection plane. Our work reveals the essential role of spin degree of freedom in the optical anomalous Hall effect.
\end{abstract}

\maketitle

%\tableofcontents
%{\it Introduction}:

Recent interest in chiral superconductors has been driven by their potential relevance to Majorana fermions and quantum computing~\cite{Kitaev2001Unpaired,Ivanov2001NonAbelian,Lutchyn2010Majorana,Asahi2012Topological,Kozii2016Three,He2021Optical,Xia2006High}. Characterized by pairing mechanisms that break time reversal symmetry, intrinsic chiral superconductivity can in principle be probed by the magneto-optical Kerr effect~(MOKE) due to the optical anomalous Hall conductivity in the absence of magnetic field~\cite{Kapitulnik2009Polar}. Such MOKE signal has been observed in Sr${}_2$RuO${}_4$~\cite{Xia2006High}, heavy fermion systems~\cite{Schemm2014Observation,Hayes2021Multicomponent,Levenson2018Polar}, high-temperature superconductors~\cite{Xia2008Polar}, and other unconventional superconductors~\cite{Farhang2023Revealing,Nandkishore2011Polar,Karapetyan2014Evidence,Saykin2023High}, suggesting them as candidates for intrinsic chiral superconductors. However, the mechanism underlying the Kerr signal remains a subject of ongoing debate. Due to the Galilean invariance~\cite{Read2000Paired}, a clean single-band chiral superconductor cannot support a nonzero Kerr signal~\cite{Lutchyn2010Majorana}. One route to breaking the Galilean invariance is through adding orbital degree of freedom to pairing, such as the inter-orbital pairing~\cite{Taylor2012Intrinsic,Wysokiifmmode2012Intrinsic}. Although they can yield a Kerr signal, previous theories~\cite{Taylor2012Intrinsic,Wysokiifmmode2012Intrinsic,Mineev2012whether,Gradhand2013Kerr} rely critically on the multi-orbital nature and the spin degree of freedom only plays the auxillary role of renormalizing the inter-orbital coupling~\cite{Taylor2012Intrinsic,Robbins2017Effect,Wang2017Intrinsic}. As a result, there exists skeptism about the existence of the interband pairing which should be small as required by Landau phase-transition theory~\cite{Mineev2012whether}. Recently, by involving the spin degree of freedom especially the spin-orbit coupling as a key element, it has been shown that the desputable interband pairing can in principle be removed in certain cases~\cite{Zhang2020Hidden,Zhang2024Quantum}. However, an accurate and complete understanding of role of the spin degree of freedom in the optical anomalous Hall effect intrinsic chiral superconductors is still missing.

In normal state of magnetic crystals, the magnetism affects the anomalous Hall conductivity through the following empirical law~\cite{Nagaosa2010Anomalous}: $\sigma_{xy}^{\rm Hall}=\sigma_H H_z+\sigma_M M_z$. This relation is deeply rooted in the Onsager's reciprocal relation~\cite{Landau2013Electrodynamics}: $\sigma_{xy}^{\rm Hall}(\bm H,\bm M)=-\sigma_{yx}^{\rm Hall}(-\bm H,-\bm M)$. It can persist to the optical response region, and hence leads to various configurations of the MOKE geometry. In intrinsic chiral superconductors, there is no local magnetism or equivalently exchange field, as the time reversal symmetry breaking occurs in the Cooper channel instead of the exchange channel. A generalized Onsager's relation is then highly relevant, which can unravel the intricate relation between the time-reversal-symmetry-breaking pairing and the anomalous Hall conductivity.

%In the normal state physics of non-superconductors, time-reversal symmetry breaking of the spin degree of freedom—typically manifested as intrinsic magnetization—can be transferred to the orbital degrees of freedom via spin-orbit coupling, giving rise to MOKE signals. This effect reflects a violation of Galilean invariance. The mechanism of MOKE is closely tied to the anomalous Hall conductivity $\sigma_H$. The properties of normal-state Hall dynamic conductivity can be categorized into three distinct structural frameworks: (i) Kinetic coefficients symmetric structure: The symmetry properties of the kinetic coefficients are governed by Onsager's reciprocity relations~\cite{Landau2013Electrodynamics,Wang2014Polar}, $\sigma_{xy}(\omega,\boldsymbol{M})=\sigma_{yx}(\omega,-\boldsymbol{M})$. (ii) Macroscopic structure: In the Landau paradigm~\cite{Landau2013Electrodynamics, Gong2017Discovery,Nagaosa2010Anomalous}, the exchange field $\boldsymbol{M}$ serves as the order parameter, and the antisymmetric part of the Hall conductivity is an odd function of the magnetization, satisfying $\sigma_{H}(\omega,\boldsymbol{M})=-\sigma_{H}(\omega,-\boldsymbol{M})$; . The two frameworks reflect the thermodynamic consequences of the system's microscopic time-reversal symmetry.

In this Letter, by employing the time reversal symmetry and gauge symmetry, we first establish generalized Onsager's relations for the optical anomalous Hall conductivity in single-orbital and spinful systems. We then perform the down-folding procedure to obtain an effective electronic Hamiltonian. Interestingly, the pairing leads to a momentum-space magnetism, which then unifies the mechanism of the optical anomalous Hall effect in magnetic materials and chiral superconductors as from spin-splitting in the electron Hamiltonian. Specifically, in chiral superconductors, such momentum-space magnetism has two different origins. The first one is from the non-unitary pairing, which also leads to a net angular momentum of the Cooper pair. The remaining one is due to the joint effect of the spin-orbit coupling and unitary paring, which has been largely overlooked previously. 

Such momentum-space magnetism can be either ferromagnetic or antiferromagnetic.  Using concrete examples, we demonstrate their unique features. Moreover, by exploring the magnetism from unitary pairing, we find that it can yield a MOKE signal beyond the scope of the empirical law, i.e., an orthogonal MOKE signal which has the set-up of the transverse MOKE and the magnetism-dependence of the polar MOKE. This is thus the superconductor version of the in-plane anomalous Hall effect observed recently.

\textit{Generalized Onsager's relation.} To study the role of the spin degree of freedom,  we consider a spinful system with a single orbital.  The corresponding Bogoliubov-de Gennes (BdG) Hamiltonian reads as $\mathcal{H}_{\text{BdG}}=\sum_{\boldsymbol{k}}\frac{1}{2}\boldsymbol{\Psi}^{\dagger}\hat{H}_{\rm BdG}(\bm k)\boldsymbol{\Psi}$,  where the Nambu spinor $\boldsymbol{\Psi}=\left(c_{k\uparrow},c_{k\downarrow},c_{-k\uparrow}^{\dagger},c_{-k\downarrow}^{\dagger}\right)^T$ and
\begin{equation}
\label{BdG Ham}
\hat{H}_{\rm BdG}(\bm k)=\left(\begin{array}{cc} \hat{H}_e(\bm k) & \hat{\Delta}(\bm k) \\ \hat{\Delta}^\dagger (\bm k) & \hat{H}_h(\bm k) \end{array}\right),
\end{equation}
$\hat{H}_e(\bm k)=A_0(\bm k)\sigma_0+{\bm A}({\bm k})\cdot{\bm \sigma}$ with $\bm \sigma$ being spin Pauli matrices, and $\hat{H}_h({\bm k})=-H_e^T(-{\bm k})$ due to the particle-hole symmetry. We comment that $H^e$ preserves the time-reversal symmetry. The pairing part takes the following form~\cite{Balian1963Superconductivity}: $\hat{\Delta}(\bm k)=\left[d_0(\bm k)+{\bm d}(\bm k)\cdot{\bm \sigma}\right]i\sigma_y.$

The MOKE signal is determined by the optical anomalous Hall conductivity $\sigma_{xy}^H$. For illustration purpose, we have restricted the Hall-deflection plane to be the $xy$ plane and the general case can be obtained strightforwardly.
Within Kubo's linear response theory, we have 
\begin{equation}
\sigma^H(\omega)\equiv -\frac{1}{2i\omega}\left[\Pi_{xy}(\omega+i0^+)-\Pi_{yx}(\omega+i0^+)\right],
\end{equation}
where 
\begin{align} \label{eq_pi}
\Pi_{xy}(i\omega_n)=\frac{e^2k_BT}{2\hbar^2}\sum_{m\bm k}\Tr\left[\hat{J}_x\hat{G}(\boldsymbol{k},i\Omega_m)\hat{J}_y\hat{G}(\boldsymbol{k},i\Omega_m+i\omega_n)\right]\,,
\end{align}
$\Omega_m$ and $\omega_n$ are the fermionic and bosonic Matsubara frequencies, respectively, $\hat{J}_x$ and $\hat{J}_y$ are current operators, and $\hat{G}(\bm k,i\omega_n)=[i\omega_n-\hat{H}_{\rm BdG}(\bm k)]^{-1}$ denotes the Gor’kov Green’s function.

We can then analyze the structure of the optical anomalous Hall conductivity. We first employ the time reversal operation $\mathcal{T}$. Since our system is spinful, $\mathcal{T}=i\sigma_y K$. Using the fact that $\hat{H}_e({\boldsymbol{k}})$ is time-reversal symmetric and that the singlet and triplet pairing are $\bm k$-even and $\bm k$-odd, respectrively, we find that $\mathcal{T}^{-1}\hat{H}_{\rm BdG}(\bm k; \Delta(\bm k))\mathcal{T}=\hat{H}^{\rm BdG}(-\bm k; \Delta^*(\bm k))$. By inserting $\mathcal{T}\mathcal{T}^{-1}$ in the trace part of Eq.~\eqref{eq_pi} and shifting $\mathcal{T}$ to the end, we find the following Onsager's relation~\cite{SI}
\begin{equation}\label{eq_first}
\sigma^H_{xy}(\omega;\hat{\Delta}(\boldsymbol{k}))=-\sigma^H_{xy}(\omega;\hat{\Delta}^*(\boldsymbol{k}))\,.
\end{equation}
Here $\sigma^H_{xy}(\omega;\hat{\Delta}(\boldsymbol{k}))$ represents the functional dependence of $\sigma^H_{xy}$ on the form of pairing $\Delta(\bm k)$. Equation.~\eqref{eq_first} demonstrates that the imaginary part of $\Delta$ flips sign under the time reversal operation~\cite{Sigrist239Phenomenological} and hence plays the role of order parameter associated with the time reversal symmetry.

Besides the time reversal symmetry, $\sigma^H_{xy}(\omega;\hat{\Delta}(\boldsymbol{k}))$ should also be constrained by the gauge symmetry of the pairing potential~\cite{Nambu1960Quasi}. Consider a rotation $\mathcal{S}=\mathrm{e}^{-i\frac{\phi}{2}\tau_z}$, which acts purely on the particle-hole degree of freedom. It changes the Hamiltonian in the following way: $\mathcal{S}^{-1}\hat{H}_{\rm BdG}(\bm k; \Delta(\bm k))\mathcal{S}=\hat{H}_{\rm BdG}(-\bm k; \Delta(\bm k)e^{i\phi})$. Since the current operator is always diagonal in the particle-hole degree of freedom, it is unchanged by $\mathcal{S}$, i.e., $\mathcal{S}^{-1}\hat{J}_{x,y}\mathcal{S}=\hat{J}_{x,y}$. Using similar methods, we find that 
\begin{equation}\label{eq_gauge}
\sigma_{xy}^{H}(\omega, \hat{\Delta}(\bm k))=\sigma_{xy}^{H}(\omega, \hat{\Delta}(\bm k)\mathrm{e}^{i\phi})
\end{equation}
This indicates that the optical Hall conductivity should be invariant by adding a global phase factor to the pairing Hamiltonian. 

The above two constraints leads to the following feature of $\sigma_{xy}^H$. It should always depends on the combined factor $d_i(\bm k)$ and $d_j^*(\bm k)$ to satisfy Eq.~\eqref{eq_gauge}. Moreover, there should be at least one critical combined factor that is purely imaginary, and $\sigma_{xy}^H$ will be an odd function of such factor, to fulfill Eq.~\eqref{eq_first}. Based on our prevous analysis, such purely imaginary factor flips sign under the time reversal operation and hence can play the role of the effective magnetic order parameter.

%Although our disucssion up to now is for the spin degree of freedom in the BdG Hamiltonian, it can be extended to discuss the pseudospin degree of freedom, from, e.g., differnt orbitals. In this case, Eq.~\eqref{eq_gauge} stays the same. For the Onsager's relation, the time reversal operator takes the form $\mathcal{T}=K$. Interestingly, its effect on the Hamiltonian stays the same, i.e., $\mathcal{T}^{-1}\hat{H}^{BdG}(\bm k; \Delta(\bm k))\mathcal{T}=\hat{H}^{BdG}(-\bm k; \Delta^*(\bm k))$. Therefore, the Onsager's relation does not change as well. Since it is proposed that the MOKE signal can have a multi-orbital origin, this extention of structure of $\sigma_{xy}^H$ can allow a unified understanding of the intrinsic MOKE signal.

\textit{Momentum-space magnetism from pairing}. In magnetic materials, the anomalous Hall effect has a clear origin, i.e., the spin-splitting in the electron Hamiltonian from magnetism. Inspired by this, we use the down-folding method to project out the hole sector and identify the role of pairing in the remaining electron Hamiltonian. Such process is valid since the strength of pairing is usually small comparied with other energy scales in concern, such as the hopping energy and the spin-orbit coupling. The effective electron Hamiltonian reads as
\begin{align}
\hat{H}_e^{\text{eff}}(\bm k)=&\hat{H}_e(\bm k)+\hat{\Delta}(\bm k)\left[E-\hat{H}_h(\bm k)\right]^{-1}\hat{\Delta}^{\dagger}(\bm k)\,.
\end{align}
By inserting the form of $\hat{H}_e(\bm k)$ and $\hat{\Delta}(\bm k)$, we find that $H_e^{\text{eff}}(\bm k)$ contains two parts~\cite{SI}, i.e.,
\begin{align}\label{eq_eff}
H_e^{\text{eff}}(\boldsymbol{k})=\widetilde{A}_0(\bm k) \sigma_0+\widetilde{\bm A}(\bm k)\cdot \bm \sigma+\bm m(\bm k)\cdot \bm \sigma\,.
\end{align}
 In the first two terms, $\widetilde{A}_0(\bm k)$ and $\widetilde{\bm A}(\bm k)$ are even and odd function of $\bm k$, respectively. The first two terms thus respect the time-reversal symmetry, and are just $\hat{H}_e(\bm k)$ with renormalized coefficients due to pairing.

The most interesting term in Eq.~\eqref{eq_eff} is the last one. The coefficient $\bm m$ can be further divided into two parts~\cite{SI}, i.e., $\bm m(\bm k)=\bm m_{\rm NU}(\bm k)+\bm m_{\rm U}(\bm k)$, where
\begin{align}
\label{eq_mag1}
\bm m_{\rm UN}(\bm k)&=i\delta_0 [E+A_0(\bm k)] [\bm d(\bm k)\times \bm d^*(\bm k)]\,,\\
\label{eq_mag2}\bm m_{\rm U}(\bm k)&=i\delta_0 \bm A(\bm k)\times[d_0(\bm k) \bm d^*(\bm k)-d_0^*(\bm k) \bm d(\bm k)]\,,
\end{align}
and $\delta_0=1/[(E+A_0(\bm k))^2-|\bm A(\bm k)|^2]$.

 Equation~\eqref{eq_mag1} and \eqref{eq_mag2} are one of main results in this work since they unify the mechanism of optical anomalous Hall effect in magnetic materials and chiral superconductors as from spin-splitting in the electronic Hamiltonian. To see this, we first note that both $\bm m_{\rm UN}$ and $\bm m_{\rm U}$ contain purely imagninary combinations of the pairing, i.e., $\bm d(\bm k)\times \bm d^*(\bm k)$ and $d_0(\bm k)\bm d^*(\bm k)-c.c.$. According to previous Onsager's relations, they should flip sign under the time reversal operation, together with the optical anomalous Hall conductivity. Therefore, $\bm m$ can act as the effective order parameter associated with the time-reversal symmetry breaking and the anomalous Hall effect. Since it emerges in momentum space, we refer to it as the momentum-space magnetism. One can also directly check that $\bm m\cdot \bm \sigma$ flips sign under the time reversal operation, confirming the role of $\bm m$. The average of $\bm m_{\rm U}$ can also give rise to a net spin magnetization~\cite{Hu2021Spontaneous}. Here we further recognize the role of $\bm m$ in the orbital motion of Bogoliubov quasiparticles, which is guaranteed by the Onsager's relation.

The physical origin of $\bm m$ can be understood by taking the following product~\cite{Leggett1975A}
\begin{align} \label{eq_prod}
\hat{\Delta}\hat{\Delta}^\dagger=(|d_0|^2+|\bm d|^2) \hat{I}+2{\rm Re}(d_0 \bm d^*) \cdot \bm \sigma+i(\bm d\times \bm d^*)\cdot \bm \sigma\,.
\end{align}
We then find that $\bm m_{\rm UN}$ directly corresponds to the last term in Eq.\eqref{eq_prod}, which makes the pairing non-unitary. Such a term will also bring a net angular momentum to the Cooper pair. Equation \eqref{eq_eff} and~\eqref{eq_mag1} then suggest that the angular momentum of Cooper pair can manifest as a momentum space magnetism and leads to a nonzero optical anomalous Hall effect. This mechanism is quite similar to that in the magnetic crystal and the spin-orbit coupling is also needed to break the spin-group symmetry~\cite{Chen2024Enumeration}.

In comparison, $\bm m_{\rm U}$ has a different meaning. Since $\bm m_{\rm U}\propto {\rm Im}(d_0 d^*)$, the pairing is still unitary and the Cooper pair does not carry a net angular momentum. Equation~\eqref{eq_mag2} then suggests that such a pairing can generate a mmentum space magnetism only when the pairing is a mixed $\bm k$-even and $\bm k$-odd pairing, and the spin-orbit vector $\bm A(\bm k)$ is perpendicular to the pairing vector $\bm d(\bm k)$.

Due to the Fermi surface instability, it has been proposed that the spin-orbit coupling and unconventional magnetism can be dynamically generated in the momentum space via the number-conserving channel in the Coulomb interaction~\cite{Wu2004Dynamic,Qian2025Fragile}. In sharp contrast, superconductivity occurs in the Cooper channel which is not number-conserving. Our result in Eq.~\eqref{eq_mag1} then states that by projecting out the hole sector, chiral pairing together with the spin-orbit coupling can induce a magnetism resembling previously studied unconventional magnetism. 

After recognizing the momentum-space magnetism, one can then derive the structure of the optical anomalous Hall conductivity. According to the Onsager's relation in Eq.~\eqref{eq_first}, the optical anomalous Hall conductivity should be an odd function of $\bm d\times \bm d^\star$ and ${\rm Im}(d_0\bm d^\star)$, and hence $\bm m$ and the induced spin magnetization $\bm M$. Due to the relative smallness of the pairing strength, we can then expand the optical anomalous Hall conductivity with respect to $\bm M$, i.e.,
\begin{align}\label{eq_sigmam}
\sigma_{ij}=A_{ijk}^{(1)}M_k+A_{ijkk_1k_2}^{(3)} M_k M_{k_1}M_{k_2}+\cdots\,.
\end{align}
We emphasize that such relation is induced by the Onsager's relation and hence only works when $\bm M$ is tuned by the pairing potential.

\begin{figure}[t]
	\centering
	\includegraphics[width=0.48\textwidth]{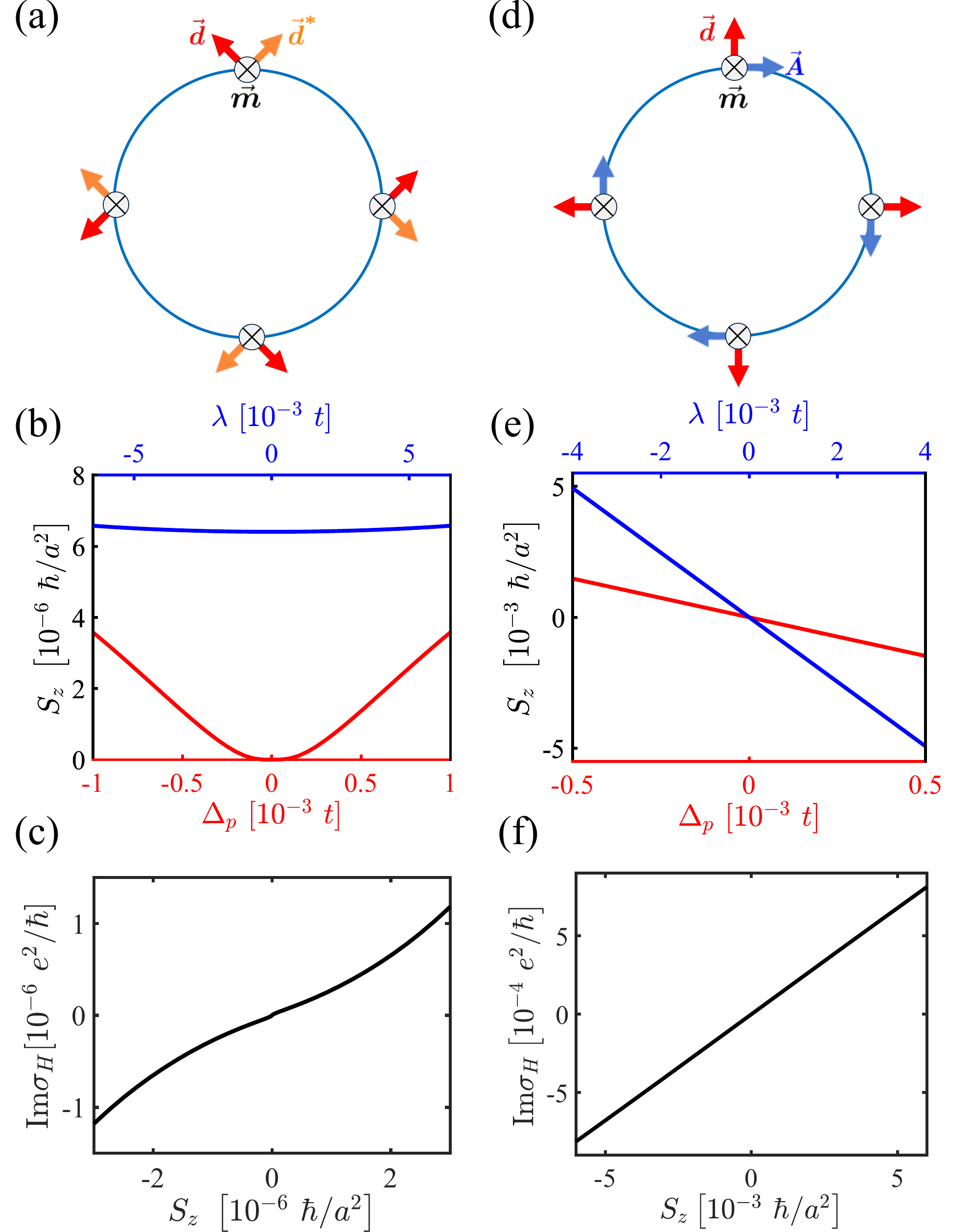}
	\caption{Momentum-space Ferromagnetism from non-unitary (a) and unitary (d) pairing mechanisms. The resulting spin angular momentum as a function of spin-orbit coupling and pairing strength from these two mechanisms are shown in (b) and (e), respectively. In both cases, the optical anomalous Hall conductivity are subject to the constraint in Eq.~\eqref{eq_sigmam}, as shown in (c) and (f). Here $t$ is the nearest neighbor hopping strength.}
	\label{fig-FM}
\end{figure}

{\it Ferromagnetism and non-collinear antiferromagnetism in momentum space.}---The momentum-space magnetism from Eq.~\eqref{eq_mag1} and \eqref{eq_mag2} can be of different type. The most common one is the ferromagnetism. It can be easily generalized in the non-unitary pairing case when the Cooper pair carries a net angular momentum, i.e., when $\bm d$ is different from $\bm d^*$ as shown in Fig.~\ref{fig-FM}(a). For example, in ${\rm Sr_2RuO_4}$ with $D_{4h}$ point-group symmetry, there can be two different p-wave channels in the orbital basis. In each channel, the spin degree of freedom can be added, and a possible pairing potential then has the form $d_0=0$ and $\bm d=(\Delta_1 k_x+i \Delta_2 k_y, \Delta_1 k_y+i \Delta_2 k_x, 0)$ where $\Delta_1$ and $\Delta_2$ are two parameters characterizing the strength of pairing in different group representations in the orbital basis\cite{Zhang2020Hidden,Zhang2024Quantum}. It can then generate a momentum space magnetism through the non-unitary channel, which reads as 
\begin{align}
\bm m_{\rm UN}\propto(0, 0, 2\Delta_1\Delta_2 (k_x^2+k_y^2))\,.
\end{align}

Such magnetism has several features. First, the total spin magnetization should be quadratic about the strength of $p$-wave pairing. It is also quadratic about the spin-orbit coupling strength which enters $\bm m_{\rm UN}$ in the proportional factor, i.e., $\delta_0$ in Eq.~\eqref{eq_mag1}. Both of them has been confirmed by numerical calculation as shown in Fig.~\ref{fig-FM}(b). To further generate a nonzero optical anomalous Hall effect as proposed in Ref.~\cite{Zhang2024Quantum}, the spin-orbit coupling is also needed to couple the time-reversal symmetry breaking in the spin degree of freedom to the orbital motion, just like that in the normal state. The resulting optical conductivity in Fig.~\ref{fig-FM}(c) confirms the structure in Eq.~\eqref{eq_sigmam}.

Momentum-space ferromagnetism can also arise from the unitary pairing, due to the mismatch between $\bm A(\bm k)$ and $\bm d(\bm k)$ as shown in Fig.~\ref{fig-FM}(d). A possible material system is $\alpha-{\rm CaPtAs}$ which has a non-centrosymmetric tetragonal structure. Its space group is $I4_1md$~(No.109) with a $C_{4v}$ point group. Due to the lack of inversion symmetry and the participation of Pt, it can has a large Rashba-type spin-orbit coupling. Experimental signals such as the superfluid density suggests a mixed pairing scenario and the zero-field $\mu$SR further suggests a time-reversal symmetry breaking pairing~\cite{Shang2020Simulaneous,Nagase2023Observation}. A minimal model for such a case then contains a s-wave pairing and a single orbital channel p-wave pairing. In this case, the momentum-space magnetism can come from the unitary mechanism. For example, we consider the following pairing function
\begin{align}
\hat{\Delta}_{A_{1}+iA_{2}}=
\begin{pmatrix}
-\Delta_p (ik_x+k_y)  && \Delta_s\\
-\Delta_s  && \Delta_p(ik_x-k_y)
\end{pmatrix}\,.
\end{align}
Here the s-wave and p-wave pairing involve the $A_{1}$ and $A_{2}$ representation of $C_{4v}$ group with $\Delta_s$ and $\Delta_p$ being the respective strength. We further adopt a Rashba-type spin-orbit coupling in the normal state, i.e., $\bm A(\bm k)=\lambda(-k_y,k_x,0)$.
The resulting momentum-space magnetism then reads as
\begin{align}
 \bm m_{\rm U}\propto (0,0,-2\lambda \Delta_s\Delta_p(k_x^2+k_y^2))\,.
 \end{align}

In sharp contrast to the ferromagnetism from the non-unitary pairing, the total spin magnetization from the unitary pairing is an odd function of the $p$-wave pairing strength and spin-orbit coupling strength $\lambda$, as required by the time reversal symmetry and spin group symmetry $C_{2z}^s$ respectively, and also reflected in Fig.~\ref{fig-FM}(e). In according to the Onsager's relation, when the p-wave pairing potential changes, the resulting optical anomalous Hall conductivity also obeys the constraint in Eq.~\eqref{eq_sigmam}, as shown in Fig.~\ref{fig-FM}(f).

The momentum-space magnetism can have a more complicated structure, such as non-collinear antiferromagnetism. This can naturally appear in the unitary pairing case when the $k$-dependence in $\bm d$ and $\bm A$ have different angular momentums. As a concrete example, we consider a continuum model subject to $C_{3v}$ point group. The model Hamiltonian reads as
\begin{equation}
\label{warping}
H_k^e=A k^2+\lambda \hat{z}\cdot (\bm k\times \bm \sigma)+\frac{\alpha}{2}(k_+^3+k_-^3)\sigma_z\,,
\end{equation}
where the first term is the kinetic energy, the second term is the Rashba spin-orbit coupling, and the last term describes the trigonal warping conforming to the threefold rotational symmetry.  Such a model can be used to describe the hetero-structure Bi/Ni, which exhibits chiral superconducting behavior~\cite{Gong2015Possible,Zhou2017Magnetic,Chao2019Superconductivity,Cai2023Nonreciprocal,Gong2017Time}. 

\begin{figure}[t]
	\centering
	\includegraphics[width=0.48\textwidth]{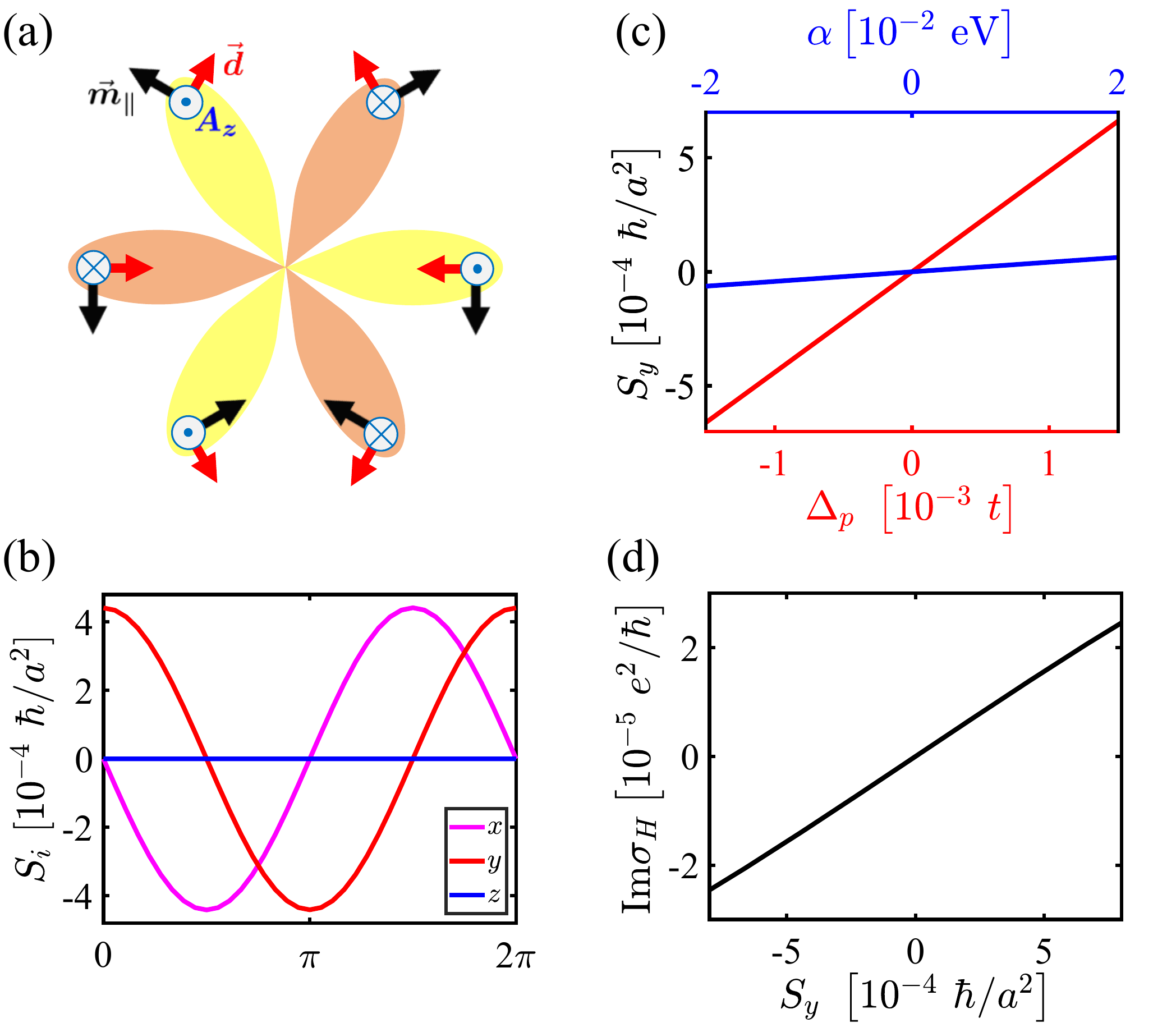}
	\caption{Momentum-space antiferromagnetism. (a) The origin of the in-plane spin texture. (b) The spin polarization $(S_x,S_y,S_z)$ as a function of $\theta$. (c) The net y-th component of spin polarization as a function of warping and p-wave pairing strength. (d) The relation between the optical anomalous Hall conductivity ans $S_y$.}
	\label{fig-PAHE}
\end{figure}

We now add chiral pairing potential. To reflect the fact that the magnetism in Ni is in-plane, we choose two-dimensional representation of $C_{3v}$ in the pairing potential. As discussed in Ref.~\cite{Gong2017Time}, the pure orbital pairing is d-wave, i.e., $d_{x^2-y^2}+id_{xy}$. When the spin degree of freedom is considered, such a d-wave pattern can come from two different vectors $\bm k$ and $\bm \sigma$, i.e., $d_{x^2-y^2}\rightarrow \phi_1=k_x\sigma_x-k_y\sigma_y$ and $d_{xy}\rightarrow \phi_2=k_x\sigma_y+k_y\sigma_x$. Momentum-space magnetism can emerge through the non-unitary mechanism by adding a phase factor between two basis functions, i.e., $\phi_1\pm i\phi_2$. However, $\bm m_{\rm UN}\parallel \hat{z}$, different from the in-plane magnetism in Ni. We then adopt the unitary mechanism and add the s-wave pairing potential. The final pairing potential reads as
\begin{align}
\hat{\Delta}=[\Delta_s\sigma_0+i\Delta_p (\phi_1 \cos\theta+\phi_2\sin\theta)]i\sigma_y\,,
\end{align}
where $\Delta_s$ and $\Delta_p$ are s- and p-wave pairing strength, and $\theta$ changes the direction of in-plane magnetism, as shown later.

Due to the mismatch between $\bm A$ and $\bm d$, the momentum-space magnetism reads as
\begin{align}
m_x&\propto \alpha \Delta_s \Delta_p  (-k_y\cos\theta+k_x\sin\theta)(k_+^3+k_-^3)\notag\\
m_y&\propto \alpha \Delta_s \Delta_p  (-k_y\sin\theta-k_x\cos\theta)(k_+^3+k_-^3) \notag\\
m_z &\propto 2\lambda \Delta_s\Delta_p[(k_x^2-k_y^2)\cos\theta+2k_xk_y\sin\theta]\,.
\end{align}
The three-fold rotational symmetry dictates an antiferromagnetism, which has a g-wave pattern for the in-plane direction as shown in Fig.~\ref{fig-PAHE}(a). With the spin-orbit coupling in the normal state, the condensate can have a net ferromagnetic moment, whose direction can be tuned by the $\theta$ angle~(Fig.~\ref{fig-PAHE}(b)) and whose magnitude is linear in the warping strength and p-wave pairing strength~(Fig.~\ref{fig-PAHE}(c)) as required by the mirror-y symmetry.

Interestingly, the resulting optical anomalous Hall effect is of in-plane nature. This can be seen from Fig.~\ref{fig-PAHE}(d), where $\sigma_{xy}$ shows linear dependence on $S_y$ instead of $S_z$. Symmetry-wise speaking, the anomalous Hall effect requires the breaking of vertical mirror symmetry, while the normal state respects the $C_{3v}$ symmetry, where $\mathcal{M}_x$ is preserved. Such mirror symmetry is further broken by the chiral pairing, rendering the linear dependence between $\sigma_{xy}$ and $M_y$. Our result then represents a striking analog of the in-plane anomalous Hall effect in normal magnetic metals~\cite{Zhou2022Heterodimensional,Cao2023In,peng2024observation,Wang2023Absence}, which explore the full tensorial strcture of the coefficient $A_{ijk}$ in Eq.~\eqref{eq_sigmam}. It can leads to a Kerr effect with the transverse setup but linear $\bm M$ dependence.

\textit{Acknowledgments} The authors thank Zhi Wang for valuable discussions. The authors are supported by the National Natural ScienceFoundation of China (Grant No. 12234017 and Grant No. 12374164). Q. N. is also supported by the National Key R${\rm \&}$D Program under grant Nos. 2023YFA1406300. Y.G. is also supported by the Innovation Program for Quantum Science and Technology (Grant No. 2021ZD0302802). This work was partially carried out at Instruments Center for Physical Science. University of Science and Technology of China. The supercomputing service of U.S.T.C. is gratefully acknowledged.

\bibliography{ref}

\end{document}